# Heat Capacity and transport studies of the ferromagnetic superconductor RuSr$_2$GdCu$_2$O$_8$

J.L. Tallon[1], J.W. Loram[2], G.V.M. Williams[1] and C. Bernhard[3]

[1]Industrial Research Institute, P.O. Box 31310, Lower Hutt, New Zealand.

[2]IRC in Superconductivity, Cambridge University, Cambridge CB3 0HE, U.K.

[3]Max Planck Institüt fur Festkörperforschung, D-70596 Stuttgart, Germany.

Abstract:

Resistivity, thermoelectric power and heat capacity are investigated in the ferromagnetic superconductor RuSr$_2$GdCu$_2$O$_8$ with and without Zn substitution. The thermodynamic signatures of the ordering of the Ru moments and the paramagnetic Gd moments as well as the onset of superconductivity are all clearly seen and quantified. The materials are shown to exhibit bulk superconductivity which coexists with spatially-uniform ferromagnetism. They appear to be typical underdoped superconducting cuprates in which the pseudogap dominates normal-state transport, thermodynamic and substitutional properties. However, a field-induced increase in T$_c$ could suggest some degree of triplet pairing.

The discovery [1,2] of superconductivity ($T_c$=18-46K) coexisting with spatially uniform ferromagnetism ($T_{Curie}$=132K) in the hybrid ruthenate-cuprate material $RuSr_2GdCu_2O_8$ (Ru-1212) presents a remarkable new development in the study of competing magnetism and superconductivity. This material allows in principle the study of several competing magnetic and superconducting (SC) order parameters including antiferromagnetic (AF) ordering of both the Gd and Cu moments, ferromagnetic (FM) ordering of the $Ru^{5+}$ moments and Cooper pairing of the carriers on both the $CuO_2$ planes and $RuO_2$ planes. The pairing interactions offer a further possibility of competing symmetries, d-wave on the $CuO_2$ planes and p-wave on the $RuO_2$ planes [3].

In most other systems where FM and SC order parameters are observed to compete uniform coexistence is not found. For example, $ErRh_4B_4$ exhibits "re-entrant" superconductivity [4] with a $T_c$ onset at 8.7K and a lower critical temperature $T_{c2}$=0.9K where the onset of uniform ferromagnetism destroys the superconducting state. Between 0.9K and 1.5K FM and SC order parameters coexist but they accommodate by becoming spatially modulated [5]. Similar behaviour occurs in $HoMo_6S_8$ [6]. In the present Ru-1212 system the FM order parameter was shown to be spatially uniform by muon spin relaxation (μSR) from the appearance of a single well-defined precession frequency which fixes the magnitude of the local field [2]. This conclusion was confirmed by electron spin resonance (ESR) studies which showed the Gd-ESR line shifted uniformly by 600 gauss with no broadening on cooling into the FM state [7]. Both the muon and Gd probes indicate a spatially-uniform local field *throughout the entire sample*. This in itself is insufficient to infer microscopic coexistence of FM and SC. As the latter might be confined to filaments or granular surfaces, it remains to demonstrate bulk SC. The presence of bulk impurities (>1%) is ruled out by synchrotron x-ray diffraction [8], however superconducting impurities (<1%) distributed finely throughout grain boundaries could possibly give rise to many of the observed SC effects. Here we present resistivity, thermo-power and heat capacity data that confirm the occurrence of bulk superconductivity consistent with other high-temperature superconducting (HTS) cuprates at a similar doping state.

The samples, from the same batch used in the μSR study, were prepared by reacting stoichiometric $RuO_2$, $SrCO_3$, $Gd_2O_3$ and CuO first in flowing $N_2$ at 1010K then successively in flowing $O_2$ at 1050°C, 1055°C and 1060°C. Samples were slow-cooled to room temperature after each 10 hour reaction, reground and pelletised. This resulted in sample A. Scanning electron micrographs showed that these samples have a high porosity with a grain size of 0.5-2μm. A further set of samples were then annealed at 1060ºC in flowing oxygen for 6 days. These samples, B, were tough and dense with a grain size of 2-10μm. X-ray diffraction patterns for the two samples were indistinguishable but high-resolution electron microscopy [8] showed the presence of nanoscale twins in sample A which were substantially absent in sample B. Further samples were reacted as for sample B but with 0.5, 1, 2 and 3% Zn substituted for Cu. Resistivity and thermopower measurements were performed on rectangular bars cut from these pellets. In addition, high-precision differential heat capacity measurements were made on both samples A and B using the Zn-substituted sample as a non-superconducting reference in order to "back-off" the phonon contribution to the specific heat, $C_p$ [9].

Fig. 1(a)  shows the resistivity, $\rho(T)$, for samples A and B. The resistivity of the unannealed sample shows a semiconducting upturn starting at 100K and a sharp fall to zero at about 20K. The annealed sample, B, has a much lower resistivity, a weaker semiconducting upturn and a higher $T_c(R=0)$ value of about 36K. Both samples exhibit a small downturn in $\rho(T)$ at $T_M$ which McCrone et al [10] conclude reveals a small amount of spin scattering arising from exchange interaction between the moments and carriers in the $RuO_2$ planes in a scenario of itinerant ferromagnetism. As shown by Pickett et al [11] the exchange interaction with carriers on the $CuO_2$ planes is weak because the Ru moments arise in the $t_{2g}$ band which couples only to the $p_x$ and $p_y$ orbitals of the apical oxygen but not to Cu $d_{x2-y2}$ or s orbitals. The shape of $\rho(T)$ shown in Fig. 1(a) is characteristic of an underdoped cuprate and, in particular, exhibits the downward curvature at 280K indicative of the presence of the normal-state pseudogap [12]. The question arises as to whether the upward shift in $T_c(R=0)$ reflects an increase in doping state resulting from the long anneal. To address this we recall that the HTS cuprates exhibit a universal correlation between the hole concentration, p, on the $CuO_2$ planes and the thermopower, S(T) [13].

Fig. 1(b) shows the T-dependence of S(T) for the two samples A and B. The magnitude and shape is again typical of underdoped cuprates and, neglecting any contribution from carriers on the $RuO_2$ planes, the room-temperature value S(300)=73 μV/K for sample B indicates p≈0.07 per Cu. This would suggest that, were the doping state to be increased to optimal (p≈0.16), the $T_c$ value could be raised to about 100K, typical of two layer cuprates. The rather small shift in S(T) with the long anneal suggests that there is little change in p with anneal, Δp being less than 0.002. This is insufficient to explain the significant rise in $T_c(R=0)$. Moreover in both samples S(T) falls to zero at temperatures 43-45K, significantly higher than the $T_c(R=0)$ value for either sample but close to the resistive onset at 46K which is the *same* for both samples. As will be seen the thermodynamic SC transition temperature is about 46K. We conclude that these Ru-1212 samples are quite granular so that the resistivity falls sharply with the onset of superconductivity but does not reach zero until a much lower percolation temperature. The granularity in Ru-1212 may be attributed to the high density of 90° [100] rotation twins that mix the a- and c-axes [7] causing pairbreaking along the twin boundaries . This proclivity for twinning arises from the near equality of these axes (c=3.01×a). With the long anneal in samples B the twinning, and hence the granularity, is substantially reduced but not eliminated. The thermopower, in contrast to the resistivity, is much less sensitive to granularity [14] and thus falls to zero close to the thermodynamic $T_c$.

Fig. 2(a) shows $\rho(T)$ for the Zn-substituted Ru-1212 samples. Here the small fraction of Zn rapidly suppresses superconductivity, displaces the curves at first upwards in parallel according to Mathiessen's rule and causes a developing semiconducting upturn in the resistivity. The rapid suppression of SC is again typical of an underdoped cuprate reflecting the strongly-depleted density of states near the Fermi level due to the presence of the pseudogap [15]. At the same time the thermopower shown in Fig. 2(b) is little altered by Zn substitution (and certainly not in a systematic fashion) except that superconductivity is progressively suppressed. We thus infer that the doping state is not significantly altered with Zn substitution as is also the case for other HTS cuprates [16]. The 3% Zn-substituted Ru-1212 sample, in which superconductivity is completely suppressed, is therefore ideal for use as a reference sample in differential heat capacity measurements as described below.

By annealing samples at progressively higher temperatures then quenching into liquid nitrogen one may explore variation in oxygen stoichiometry in the SC and normal state (NS) properties. A series of such experiments were carried out on sample A. Curves of $\rho(T)$ for different annealing conditions are shown in the inset to Fig. 3. The resistivity is progressively increased, the semiconducting upturn enhanced and $T_c(R=0)$ is quite slowly shifted down. However, these changes are due simply to increasing granularity and the plots of $d\rho/dT$ shown in the main panel of Fig. 3 indicate that the superconducting onset at the thermodynamic transition temperature is unaltered by these anneals. The mass of the sample diminished by only 0.2mg in 1.1g reflecting a constancy of the bulk oxygen stoichiometry of 8±0.008.

The inset to Fig.4 showing the zero-field cooled (ZFC) susceptibility of sample B measured at 5.5 gauss reveals a maximal volume diamagnetic susceptibility close to 100% with a transition at 28K. This transition is lower than both the zero-resistance (36K) and thermodynamic (46K) transitions. However, closer inspection of the data on an expanded scale (see Fig. 4 main panel) shows that a diamagnetic onset coinciding with the thermodynamic transition does occur at 45K but amounts to just 0.3% of full diamagnetism. This arises because of the granularity and shows that macroscopic screening currents do not percolate at 5.5 Gauss until T<28K.

The above results reveal a superconducting system typical of the HTS cuprates with a transition temperature and normal-state transport properties in all respects consistent with granular underdoped cuprates. It seems likely therefore that Ru-1212 is a bulk superconductor like the other HTS cuprates. However, without thermodynamic measurements filamentary or surface superconductivity cannot be discounted. We have carried out differential heat capacity measurements on samples A and B relative to an equimolar 3% Zn-substituted Ru-1212 sample. Fig. 5(a) shows $\gamma_{tot} \equiv C_{tot}/T$ for applied fields of 0, 4, 9 and 13 Tesla for sample B, where $\gamma_{tot}$ includes electronic, magnetic and phonon terms. The dashed curve showing the previously measured $\gamma(T)$ for $YBa_2Cu_3O_7$ with 7% Zn substitution [9] illustrates the similarity of the phonon contribution, $\gamma_{ph}$, for the two systems. The FM transition is evident at 132 K and the low-T upturns are due to field alignment of the Gd spins. The latter is quantified in Fig. 5(b) by plotting the T-dependence of $\gamma(Gd) \approx \gamma - \gamma_{ph}$, where the phonon term is similar to that of $YBa_2Cu_3O_7$ and has an initial Debye temperature $\Theta_D(0) \approx 390K$. At high fields the curves for $\gamma_{Gd}$ agree closely with the magnetic contribution $\gamma_{mag}(T)$ for a mole fraction $f_{Gd}=1/14$ of non-interacting localised paramagnetic moments with J=7/2 and g=2, shown in the inset to Fig. 5(b). At lower fields it is necessary to include the distribution of internal fields arising from both the Gd and Ru moments in order to fit the T-dependence.

The inset to Fig. 5(a) shows that the ferromagnetic anomaly in the neighbourhood of the Curie point is not mean-field-like but exhibits strong 3D-XY fluctuations which extend to room temperature [2]. The step height, $\approx 4.2$ mJ/g-at.$K^2$, is a factor of 2 to 3 smaller than that expected for localised Ru moment ferromagnetism but would be compatible with itinerant ferromagnetism in the Ru planes if $\gamma_{Ru} = 2$-3 mJ/g.at.$K^2$ in the paramagnetic region. This is rather larger than the band structure value $\gamma_{Ru} \approx 0.8$ mJ/g.at.$K^2$ deduced from a $t_{2g}$ band width of 1.25 eV [10]. If the ferromagnetism is itinerant then the FM exchange energy must be a significant fraction of the Fermi energy [16] and both must be small ($\leq 1/3$ that for Ni).

Fig. 6 (a) and (b) show $\gamma$ relative to the reference, $\gamma - \gamma_{ref}$, for samples A and B, respectively. In both cases the SC transition sets in at about 46K and for the annealed sample has a jump $\Delta\gamma\approx0.26$ mJ/g-at.K$^2$. Sample A has a broader and weaker transition ($\Delta\gamma\approx0.2$ mJ/g-at.K$^2$) but clearly has the same onset at 46K. This confirms our conclusion that the thermodynamic transition occurs at 46K in both samples and that the establishment of macroscopic supercurrents is dominated by the granularity of the samples. The magnitude of the jump $\Delta\gamma$ (and hence the condensation energy) is small but typical of other underdoped HTS cuprates (see dotted curves in (b) for two underdoped $Bi_2Sr_2CaCu_2O_8$ samples [18]) and *confirms the presence of bulk superconductivity*. In both Ru-1212 and Bi-2212 $\Delta\gamma$ is substantially reduced from its value at optimal doping due to the presence of the pseudogap which strongly attenuates $T_c$, $\gamma$, $\Delta\gamma$ and the condensation energy $U_o = {_o}^{T_c}(S_{NS} - S_{SC})dT$ due to the loss of spectral weight available for SC [19,20]. The in-field reduction in $\Delta\gamma$ (and hence the superfluid density) is, again, comparable to underdoped Bi-2212 while the absence of a field-dependent shift in $T_c$ to lower temperature implies a very large $H_{c2}$ (> 60 Tesla). In fact, $T_c$ determined by either the midpoint or maximum slope in $\gamma(T)$ appears to move up some 4.5K in both samples when the field is increased to 4 or 5 Tesla. This raises the interesting possibility of triplet pairing (which is enhanced in a magnetic field). Leaving aside the question of intrinsic p-wave pairing interactions in the $RuO_2$ layer, theoretical considerations show that proximity coupling of d-wave SC to an itinerant FM will induce a component of p-wave order parameter near the interface in each [21].

Without investigating the differential heat capacity of a series of samples extending towards the parent undoped AF insulator it is not possible to determine the absolute NS value of $\gamma_{el}(T)$. However, as shown previously [15] we can estimate $\gamma_{el}(T_c)$ from the reduction in $T_c$ due to the fraction, y, of Zn substitution. For unitary scattering with a d-wave order parameter the initial rate of reduction in $T_c$ is given by

$$dT_c/dy = 58.5 \ (\Delta_{00}/T_{c0})^{-1} \ \gamma(T_c)^{-1} \ , \tag{1}$$

where $\Delta_{00}$ and $T_{c0}$ are the T=0 SC order parameter and $T_c$ value, respectively, in the absence of impurity scattering and $\gamma$ is in units of J/mol.K$^2$. It was found that this model gave a remarkably good description of the changes in the depression of $T_c$ by impurity substitution across the under-and over-doped regions for both $Y_{0.8}Ca_{0.2}Ba_2Cu_3O_{7-\delta}$ and $La_{2-x}Sr_xCuO_4$ using only the experimentally-observed values of $\gamma(T_c)$. The inset to Fig. 2(b) shows $T_c$ plotted as a function of Zn content for the resistive onset (▲) and mid-point (▼) as well as the thermopower onset (   ). The solid lines are the Abrikosov-Gorkov equation [15]. From the initial slope $dT_c/dy = 15$ K/% we find $\gamma=0.69$ mJ/g.at.K$^2$. This is just over one third that found for overdoped $Y_{0.8}Ca_{0.2}Ba_2Cu_3O_{7-\delta}$ [20] and about one half that found for overdoped $Bi_2Sr_2CaCu_2O_8$ [18]. The low value of $\gamma(T_c)$ and hence the high value of $dT_c/dy$ reflects, as we have noted, the presence of the pseudogap in Ru-1212.

In conclusion, from transport and heat capacity studies we have shown that Ru-1212 behaves as a typical short-coherence-length underdoped cuprate which, though granular, exhibits *bulk* superconductivity with a jump in $\gamma(T)$ at $T_c$ comparable to that observed in other HTS cuprates at the same doping state. We are therefore able to confirm the occurrence of spatially-uniform ferromagnetism coexisting with *bulk* superconductivity. This material clearly exhibits a number of features arising from the presence of the normal-state pseudogap observed in other underdoped cuprates. Although the magnetism in the $RuO_2$ layers is itinerant [11] the transport properties appear to be dominated by the $CuO_2$ planes. The upward shift in $T_c$ with field suggests some degree of triplet pairing.

**Figure Captions:**

Fig. 1 (a) T-dependence of the resistivity for as-prepared (A) and 6-day annealed (B) samples of Ru-1212. (b) T-dependence of the thermopower for the same samples.

Fig. 2. (a) T-dependence of the resistivity for annealed samples of Ru-1212 with a range of Zn concentrations (indicated in %). (b) T-dependence of the thermopower (TEP) for the same samples. Inset: Zn concentration dependence of $T_c$ from TEP onset ($\bullet$), resistive onset ($\blacktriangle$) and resistive mid-point ($\blacktriangledown$). Curves show the Abrikosov-Gorkov relation.

Fig. 3. The T-dependence of the derivative of the resistivity for Ru-1212 after annealing and quenching in oxygen at 250, 440, 570 and 660°C. Inset resistivity for the same samples.

Fig. 4. The T-dependence of the ZFC volume susceptibility for Ru-1212. Inset: the complete transition. Main panel: enlargement near $T_c$ showing the onset of the thermodynamic transition at 45K. The dotted curve shows the extrapolated NS behaviour.

Fig. 5. (a) $\gamma_{tot} \equiv C_{tot}/T$ for annealed Ru-1212. Solid curve: 0 Tesla. Light curves: 4, 9 and 13 Tesla. Dashed curve: $\gamma_{tot}$ for 7% Zn-substituted $YBa_2Cu_3O_7$ as an approximation for the phonon contribution, $\gamma_{ph}$, for Ru-1212. Inset $\Delta\gamma = \gamma - \gamma_{ph}$ for the as-prepared (A), annealed (B) and reference (ref) samples showing the anomaly at the Curie temperature. (b) $\gamma_{Gd} = \gamma - \gamma_{ph}$ at low temperature showing the ordering of the Gd moments in applied fields of 0, 4, 9 and 13 Tesla. Inset: calculated $\gamma(T)$ for non-interacting paramagnetic moments with J=7/2.

Fig. 6. Differential $\gamma(T)$ relative to the Zn-substituted reference for (a) as-prepared sample and (b) 6-day annealed sample in magnetic fields as annotated in units of Tesla. The dotted curves in (b) show $\gamma - \gamma_{ref}$ at zero field for two underdoped Bi-2212 samples.

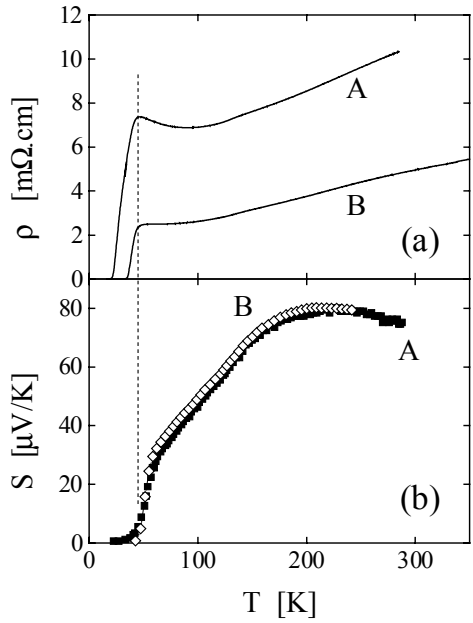

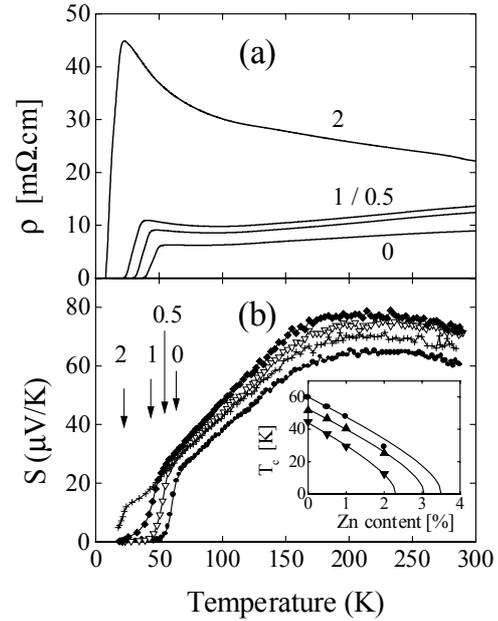

Fig. 1 (a) T-dependence of the resistivity for as-prepared (A) and 6-day annealed (B) samples of Ru-1212. (b) T-dependence of the thermopower for the same samples.

Fig. 2. (a) T-dependence of the resistivity for annealed samples of Ru-1212 with a range of Zn concentrations (indicated in %). (b) T-dependence of the thermopower (TEP) for the same samples. Inset: Zn concentration dependence of $T_c$ from TEP onset (●), resistive onset (▲) and resistive mid-point (▼). Curves show the Abrikosov-Gorkov relation.

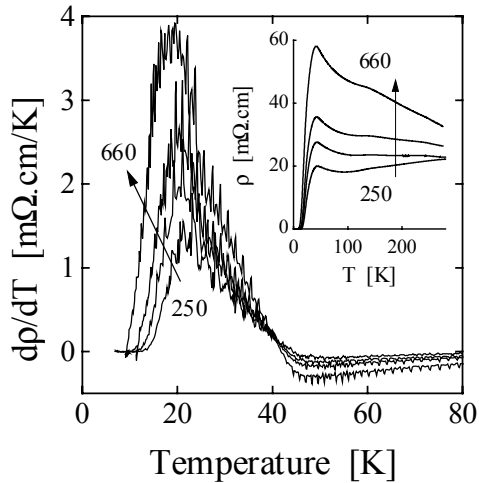

Fig. 3. The T-dependence of the derivative of the resistivity for Ru-1212 after annealing and quenching in oxygen at 250, 440, 570 and 660°C. Inset resistivity for the same samples.

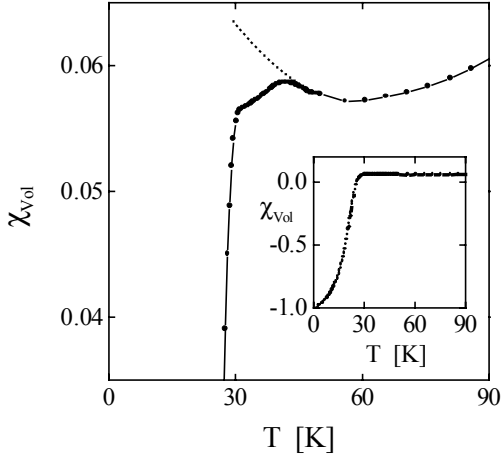

Fig. 4. The T-dependence of the ZFC volume susceptibility for Ru-1212. Inset: the complete transition. Main panel: enlargement near $T_c$ showing the onset of the thermodynamic transition at 45K.

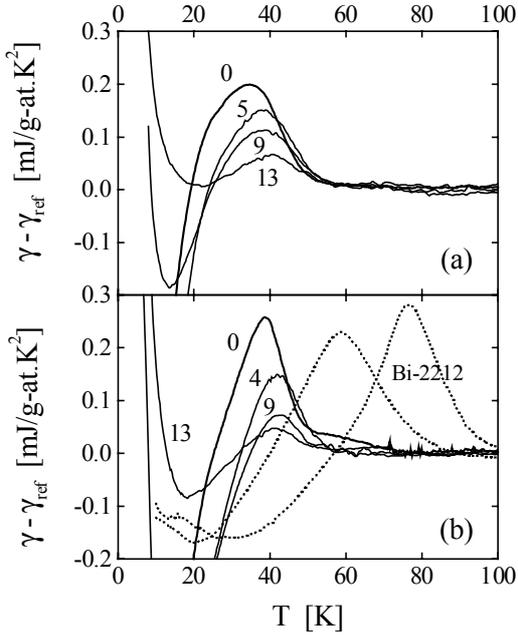

Fig. 6. Differential $\gamma(T)$ relative to the Zn-substituted reference for (a) sample A and (b) sample B in magnetic fields as annotated in units of Tesla. The dotted curves in (b) show $\gamma$-$\gamma_{ref}$ for underdoped Bi-2212.

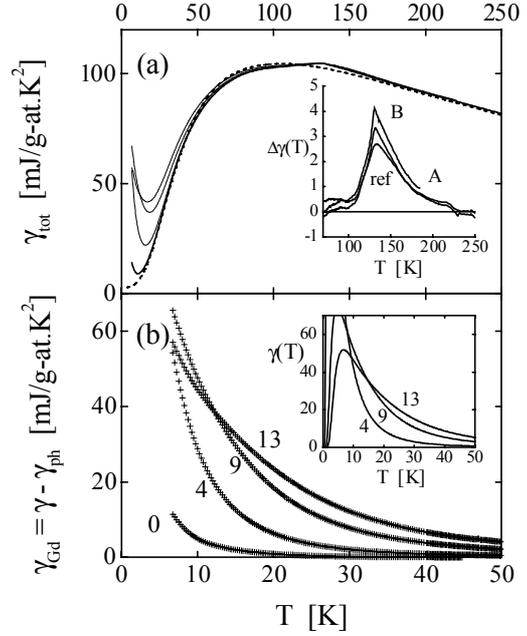

Fig. 5. (a) the total $\gamma \equiv C_p/T$ for annealed Ru-1212. Solid curve: 0 Tesla. Light curves: 4, 9 and 13 Tesla. Dashed curve: $\gamma_{tot}$ for 7% Zn-substituted YBa$_2$Cu$_3$O$_7$ as an approximation for the phonon contribution, $\gamma_{ph}$, for Ru-1212. Inset $\Delta\gamma = \gamma$-$\gamma_{ph}$ for the as-prepared (A), annealed (B) and reference (ref) samples showing the anomaly at the Curie temperature. (b) $\gamma_{Gd} = \gamma$-$\gamma_{ph}$ at low temperature showing the ordering of the Gd moments in applied fields of 0, 4, 9 and 13 Tesla. Inset: calculated $\gamma(T)$ for non-interacting paramagnetic moments with J=7/2.